\documentclass[conference]{IEEEtran}
\usepackage{array}
\usepackage{caption}
\usepackage{graphicx}
\usepackage{tabularx}
\usepackage{adjustbox}
\usepackage{booktabs}
\usepackage{multirow}
\usepackage[table,xcdraw]{xcolor}
\usepackage{hyperref}
\usepackage{amsmath}
\usepackage{subfigure}\usepackage{graphicx}
\usepackage{subfigure}
\usepackage{hyperref}

\ifCLASSINFOpdf
\else
\fi

\begin{document}
\title{Leveraging Self-Supervised Models for Automatic Whispered Speech Recognition}

\author{\IEEEauthorblockN{Aref Farhadipour}
\IEEEauthorblockA{Department of Computational Linguistics \\
University of Zurich\\
Zurich, Switzerland.\\
aref.farhadipour@uzh.ch}
\and
\IEEEauthorblockN{Homa Asadi}
\IEEEauthorblockA{Department of Linguistics\\
University of Isfahan\\
Isfahan, Iran.\\
h.asadi@fgn.ui.ac.ir}
\and
\IEEEauthorblockN{Volker Dellwo}
\IEEEauthorblockA{Department of Computational Linguistics\\
University of Zurich\\
Zurich, Switzerland.\\
volker.dellwo@uzh.ch}}

\maketitle

\begin{abstract}
In automatic speech recognition, any factor that alters the acoustic properties of speech can pose a challenge to the system's performance. This paper presents a novel approach for automatic whispered speech recognition in the Irish dialect using the self-supervised WavLM model. Conventional automatic speech recognition systems often fail to accurately recognise whispered speech due to its distinct acoustic properties and the scarcity of relevant training data. To address this challenge, we utilized a pre-trained WavLM model, fine-tuned with a combination of whispered and normal speech data from the wTIMIT and CHAINS datasets, which include the English language in Singaporean and Irish dialects, respectively. Our baseline evaluation with the OpenAI Whisper model highlighted its limitations, achieving a Word Error Rate (WER) of 18.8\% and a Character Error Rate (CER) of 4.24\% on whispered speech. In contrast, the proposed WavLM-based system significantly improved performance, achieving a WER of 9.22\% and a CER of 2.59\%. These results demonstrate the efficacy of our approach in recognising whispered speech and underscore the importance of tailored acoustic modeling for robust automatic speech recognition systems. This study provides valuable insights into developing effective automatic speech recognition solutions for challenging speech affected by whisper and dialect. The source codes for this paper are freely available \footnote{\href{https://github.com/areffarhadi/Whisper_fine_tuning_ASR}{GitHub Repository for Whisper transcription.}}\footnote{\href{https://github.com/areffarhadi/PaddleSpeech}{GitHub Repository for WavLM fine-tuning.}}.
\end{abstract}

%\begin{keywords}


\IEEEpeerreviewmaketitle

\section{Introduction}
\label{sec:intro}

Recent advancements in self-supervised models have significantly impacted various speech processing domains, such as speech recognition, speaker recognition \cite{farhadi2024comp, farhadipour2024analysis}, and speaker diarization \cite{yang2021superb}. These models, especially in English, have substantially improved system efficiency by leveraging large multilingual datasets, primarily featuring normal speech. However, these models struggle when processing atypical forms of speech, including whispered, reverberant \cite{nam2025coherence}, and disordered speech \cite{far2018dys, farh2024gamm}, as such variants are underrepresented in the training data.

Whispered speech, a distinct mode of communication, is naturally produced by most individuals for privacy or when vocal cords are compromised. Its acoustic characteristics differ markedly from normal speech, which poses a unique challenge for traditional speech recognition systems. These challenges highlight the importance of developing specialized automatic speech recognition (ASR) systems to handle whispered speech, especially considering its practical applications in healthcare and privacy-sensitive environments.

Dialectal variations, in addition to whispered speech, further complicate speech recognition. Dialects introduce significant changes in acoustic parameters, which reduce the accuracy of ASR systems trained predominantly on standard dialects. In particular, whispered speech in non-standard dialects, such as the Irish dialect, presents an even more significant challenge. No multi-modal dataset exists for whispered speech in any dialect, further compounding the problem.

This research addresses these critical gaps by focusing on the original and novel goal of developing an ASR system specifically designed to recognize whispered speech in the Irish dialect. Leveraging the capabilities of the self-supervised WavLM model, which has shown superior performance across various tasks, this study aims to enhance the system's ability to identify whispered speech accurately. By incorporating dialectal variations into the model's training, this research aims to solve a largely underexplored area.

To further emphasize the limitations of traditional models, we will utilize OpenAI’s \textit{Whisper} model \cite{radford2023robust}, a state-of-the-art system renowned for its performance in standard speech recognition tasks. Comparing \textit{Whisper} to WavLM in the context of whispered speech highlights the current limitations and underscores the necessity of specialized models for whispered and dialectal speech.

The development of an effective ASR system requires a comprehensive and diverse dataset. However, research on whispered speech remains limited due to a scarcity of high-quality data. Existing datasets, such as wTIMIT \cite{lim2010wtimit} and CHAINS \cite{grimaldi2008speaker}, primarily focus on English and fail to address the nuances of whispered speech in other languages or dialects. Furthermore, while data augmentation techniques like generating reverberant speech \cite{ko2017study} have proven successful for normal speech, they are inadequate for whispered speech conversion. Efforts to artificially convert normal speech to whispered speech remain problematic due to the unnatural and noisy outputs \cite{avdeeva2023streaming}.

Given these challenges, fine-tuning pre-trained self-supervised models offers a practical solution to the scarcity of whispered speech data. This method, as supported by recent research \cite{lin2023improving}, allows models to leverage existing knowledge and adapt to new tasks, offering a more viable alternative than traditional end-to-end (E2E) systems based on conformers or transformers. In this study, we aim to address the data scarcity challenge and demonstrate the effectiveness of fine-tuning techniques in developing robust ASR systems for whispered speech.

Previous work has explored various methods for recognising whispered speech, considering its unique characteristics and the challenge of limited data availability. As one of the first works in whispered speech recognition, \cite{ito2005analysis}, designed a system based on hidden Markov model and mel frequency cepstral coefficient features. Some works were also focused on improving the features \cite{yang2012noise, pour2014gammatonegram} and adapting the acoustic model \cite{lim2011computational,lee2014whispered, mathur2012significance}. However, Several works focused on improving pre-processing \cite{ghaffarzadegan2014ut,galic2014htk} to enhance system performance for whispered speech. 

Some studies have addressed the challenge of limited data for whispered speech by employing generative methods and speech conversion techniques. These approaches include recurrent neural networks, Gaussian mixture models, sequence-to-sequence models,  and various generative architectures \cite{toda2005nam, lian2019whisper, meenakshi2018whispered}. For example, in \cite{wagner2023generative}, the authors proposed approaches based on MelGAN and VQ-VAE, adapted explicitly for whispered speech conversion.\cite{avdeeva2023streaming} focused on low-delayed whisper-to-speech voice conversion, which can be helpful in real-life communication of people. 

In \cite{chang2021end}, an E2E method was proposed that included a frequency-weighted SpecAugment policy and a frequency-divided Convolutional Neural Network (CNN) feature extractor to better capture the high-frequency structures of whispered speech. This approach resulted in a 19.8\% reduction in Phoneme Error Rate (PER) and 44.4\% in Character Error Rate (CER) on the wTIMIT corpus. In \cite{lin2023improving}, speakers were divided into four categories based on accent and speech type, and the E2E method with the ESPnet toolbox \cite{watanabe2018espnet} was used. The study found a WER of 36.3\% for whispered speech, using pitch and filterbank features without a language model. 

Gudepu et al., \cite{gudepu2020whisper}, generated a synthetic whisper corpus from normal speech using CycleGAN, which was then used to augment data for ASR systems. This approach achieved a 23\% relative reduction in WER on whispered speech compared to the baseline. Ghaffarzadegan et al., \cite{ghaffarzadegan2016generative}, introduced two strategies to generate pseudo-whisper utterances using vector Taylor series and denoising autoencoders. These strategies significantly reduced WER for a TIMIT-trained speech recogniser on whisper recognition tasks. 

In \cite{grozdic2017whispered}, a new framework using deep denoising autoencoders and Teager-energy-based cepstral features was proposed. This framework improved whisper recognition accuracy by 31\% over traditional methods, achieving a 92.81\% word recognition rate in mismatched train/test scenarios.

The remainder of this article is structured as follows: Section \ref{sec:whsp} discusses whispered speech and its phonetic differences from normal speech. Section \ref{sec:methd} introduces the datasets and the proposed system. The results and findings are presented in Section \ref{sec:exp}, and the conclusion is provided in Section \ref{sec:concl}.

\section{Whispered speech}
\label{sec:whsp}
Whispered speech is a natural communication mode often employed in public situations to protect privacy or when one wishes to avoid being overheard. This form of speech differs significantly from normal speech in both production and acoustic properties. The fundamental difference between whispered and normal speech lies in their generation mechanisms. Whispered speech is produced by exhaling air and adjusting the shape of the pharynx so that the vocal cords do not vibrate \cite{ito2005analysis}.
While normal speech involves vocal fold vibrations to produce sound, whispered speech relies on turbulent, aperiodic airflow passing through a relatively small, nearly constant aperture formed between the vocal folds \cite{jovivcic2008acoustic,swerdlin2010effect}. 

This results in a breathy quality with less distinct harmonic patterns, making it difficult to identify formants in the spectrogram of whispered speech \cite{lin2023improving}.
In terms of production, physiological measurements using magnetic resonance imaging have shown that during the whispered speech, the supraglottal structures are constricted and shifted downward, attaching to the vocal folds to prevent their vibration \cite{tsunoda1997laryngeal}. As a result, speakers produce sounds without fully engaging their vocal cords, leading to a hushed, breathy quality. The resonance of whispered speech usually occurs at significantly higher frequencies than normally phonated speech \cite{matsuda1999acoustic}.

\begin{figure*}[t]
    \centering
    \includegraphics[width=0.9\textwidth, height=0.3\textheight]{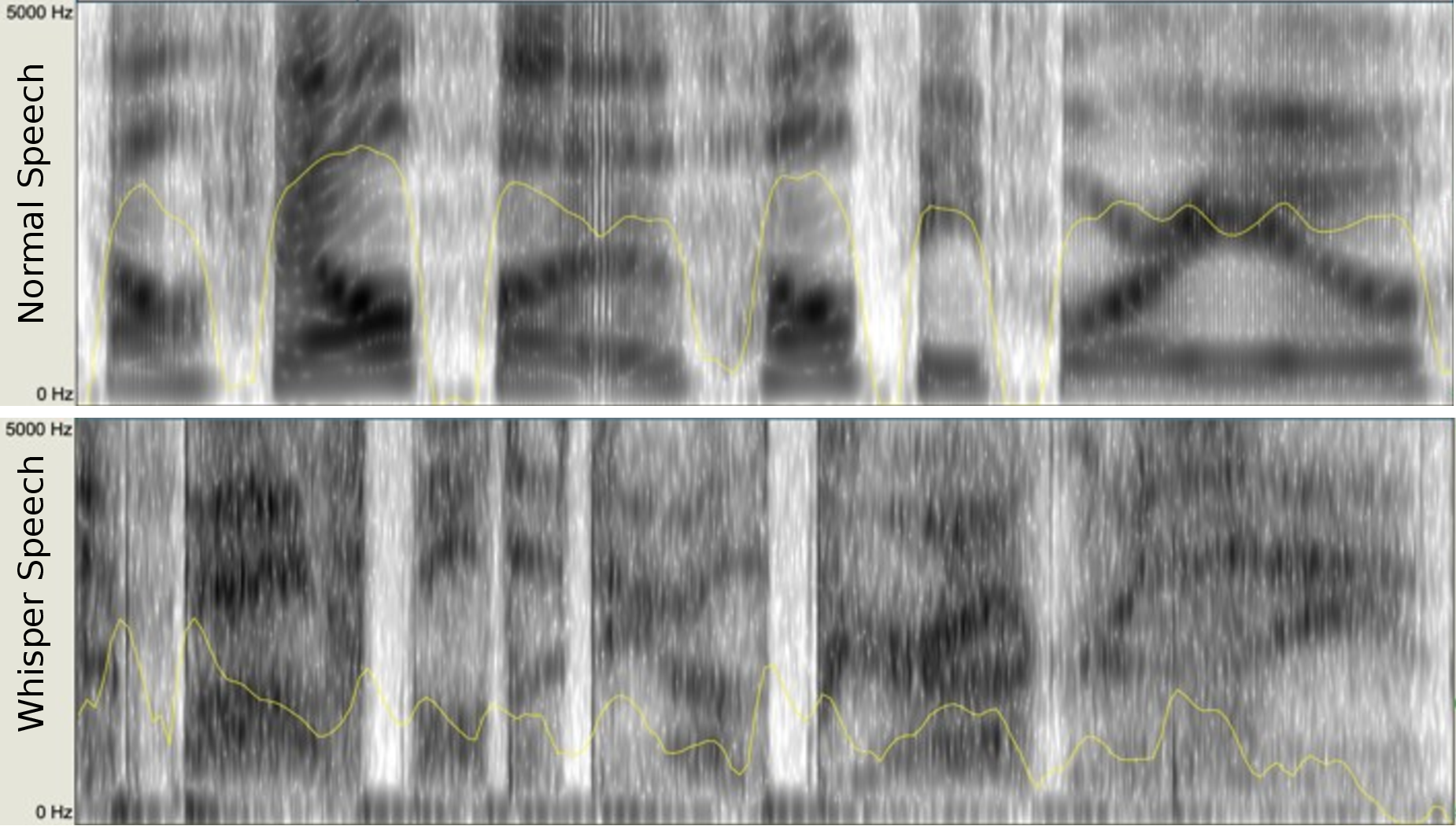}
    \caption{Spectrogram of normal and whispered speech from the same speaker}
    \label{fig:F0}
\end{figure*}
Thus, whispered sounds tend to have a higher frequency range, primarily due to increased airflow turbulence. This turbulence generates a noisier signal with less defined formant structures, which are crucial for identifying vowels and certain consonants in normal speech. Consequently, whispered speech can be less intelligible, especially in noisy environments. Furthermore, whispered speech has a lower energy contour, reduced overall amplitude, and a less distinct formant structure, particularly in the lower frequencies, along with a flatter spectral slope \cite{fan2013acoustic,matsuda1999acoustic,lin2023improving}.

The articulation of phonemes is also altered in whispered speech \cite{sharifzadeh2012comprehensive,jovivcic2008acoustic}. Research indicates that in whispered speech, the first and second formants shift to higher frequencies, and vowel duration is extended \cite{johnson2021speaker,sharifzadeh2012comprehensive,houle2020acoustic}. 

Additionally, studies have shown that the duration of consonants is prolonged compared to normal speech, with the intensity of voiceless consonants remaining unchanged. In contrast, the intensity of voiced consonants decreases \cite{schwartz1972bilabial}. 

\cite{rubin2006laryngeal} found that producing whispered speech demands greater physiological exertion than normal speech. Maintaining the necessary vocal fold position can increase tension in the laryngeal muscles, potentially leading to vocal fatigue if sustained for extended periods. These characteristics pose challenges for speech recognition systems, necessitating different approaches for analysis and synthesis compared to normal speech. 

Despite these challenges, whispered speech serves important functions in communication, often used for privacy, intimacy, or in situations where normal speech volume is inappropriate. Fig \ref{fig:F0} displays the notable differences in the spectrogram characteristics of the speech signals spoken by the same individual in both neutral and whisper modes in Praat software.

As Fig \ref{fig:F0} shows, in whispered speech, the waveform exhibits lower amplitude and lacks the regular periodicity observed in normal phonation, which indicates the absence of vocal fold vibrations. The frequency content in whispered speech is more diffuse and less defined, characterized by an evenly distributed energy spectrum and a lack of pronounced harmonic structures. The overall intensity is lower, as evidenced by the lighter grayscale in the spectrogram, and the yellow intensity contour line displays lower and less stable intensity levels compared to normal speech. 

Conversely, normal speech is characterized by higher amplitude and regular periodic patterns, with distinct harmonic bands and well-defined speech formants. The spectrogram of normal speech shows higher intensity levels, indicated by darker areas, and the yellow intensity contour line is more stable and reaches higher intensity levels, reflecting concentrated energy at specific frequencies.

\section{Methodology}
\label{sec:methd}
In this section, the WavLM model as the core of the proposed system and the CHAINS dataset are introduced as the evaluation dataset utilized in this study.

\subsection{WavLM}
The WavLM model \cite{chen2022wavlm}, a self-supervised learning (SSL) model introduced by Microsoft in 2021, aims to provide a comprehensive solution for various speech processing applications. WavLM was designed and trained for full-stack speech processing differently from other SSL models, achieving superior performance in tasks such as speaker recognition and separation, which demand an efficient understanding of acoustic models. The training process of this model involved masked speech prediction and denoising, which were performed jointly. This dual-task training strategy enables the model to predict masked speech content while learning to denoise, thereby equipping it with the ability to handle multiple speech processing scenarios beyond ASR.

\begin{figure}[t]
    \centering
    \includegraphics[width=8cm]{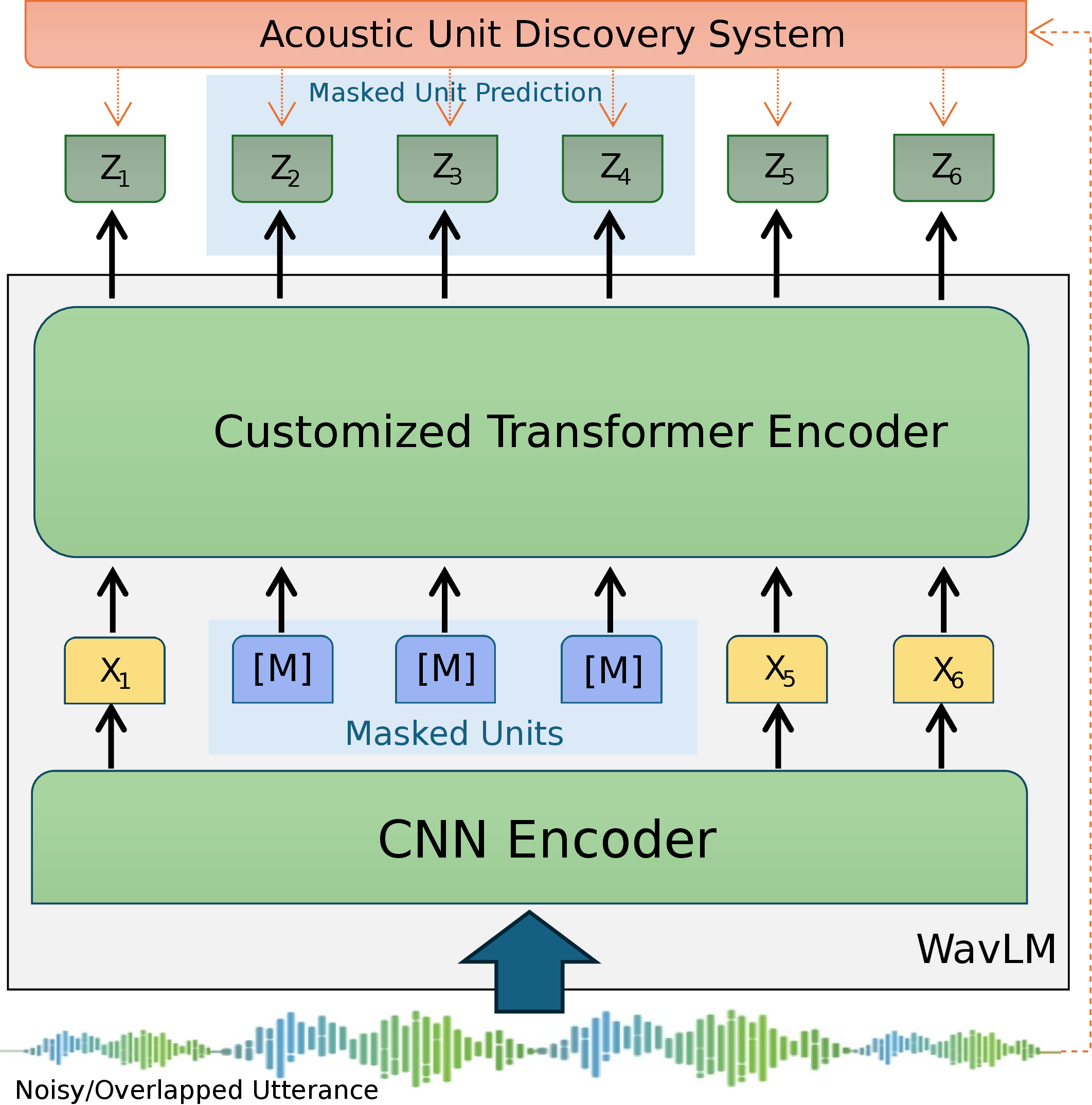}
    \caption{Architecture of WavLM model}
    \label{fig:F1}
\end{figure}
WavLM was pretrained on 94,000 hours of speech data, encompassing a variety of speech types, including read and spontaneous speech, allowing it to model a broad spectrum of speech characteristics. According to benchmarks, WavLM outperforms other SSL models across many scenarios. The model architecture incorporates a gated relative position bias in the transformer self-attention mechanism to enhance its performance. 

As depicted in Fig \ref{fig:F1}, WavLM consists of a CNN feature encoder and a transformer encoder with gated relative position bias. The CNN encoder comprises seven blocks of temporal convolution, followed by normalization and GELU activation layers, with each convolutional block representing 25 milliseconds of the input signal. The transformer section includes position embedding with a kernel size of 128 and 16 groups at its bottom \cite{chen2022wavlm}. 

By predicting the masked noisy/overlapped parts, the model learns the speaker's acoustic information from the unmasked segments and estimates the masked parts using the speaker's characteristics. This strategy significantly enhances the model's ability to learn and generalize acoustic information. 

In this network, let \( X = \{x_1, x_2, \ldots, x_T\} \) represent a speech utterance with \( T \) frames, and let the discovered hidden units be \( Z = \{z_1, z_2, \ldots, z_T\} \). Let \( M \subset [T] \) denote the set of indices to be masked for the length-\( T \) sequence \( X \). The corrupted version of \( X \), denoted as \( \hat{X} = r(X, M) \), is defined such that \( x_t \) is replaced with a mask embedding \( \hat{x} \) if \( t \in M \) \cite{hsu2021hubert}.

\subsection{Whispered Speech Dataset}
The CHAINS dataset \cite{grimaldi2008speaker} is utilized to evaluate the proposed system. This dataset includes both normal speech, referred to as \textit{solo}, and whispered speech, known as \textit{whsp}. The dataset comprises recordings from 36 speakers, dominantly with Irish dialects. To develop a model capable of accurately capturing the acoustic features of the Irish dialect, the normal speech portion of this dataset is used for system training, while the whispered speech portion is employed for evaluation. Additionally, to improve the model's performance on a whispered speech by learning its unique acoustic characteristics, the whispered and normal speech portion of the wTIMIT dataset \cite{lim2010wtimit}, which is in Singaporean dialect, is also used as auxiliary data in the training process. 

In this evaluation, only the Singaporean dialect speech from wTIMIT is used for training, ensuring that the system encounters different dialects from those used during pre-training, thereby fostering a more general acoustic model. Each speaker's evaluation data involves 33 whispered utterances from the CHAINS dataset. Further details of these datasets are presented in Table \ref{tab:t1}. 

\begin{table}[t]
\caption{Details of whisper datasets and their application as test and train data}
\label{tab:t1}
\centering
\resizebox{\linewidth}{!}{%
\renewcommand{\arraystretch}{1.5} % Increase the vertical spacing in each cell
\begin{tabular}{|l|llllll|}
\hline
Dataset & Style   & \#Utter. & \#Spkr & Dialect   & Dur. (h) & Usage \\ \hline
wTIMIT  & whisper & 8460     & 20     & Singaporean & 11       & train \\
\rowcolor[HTML]{D9D9D9} 
wTIMIT  & normal  & 4652     & 11     & Singaporean & 6        & train \\
CHAINS  & normal  & 1332     & 36     & Irish     & 3        & train \\
\rowcolor[HTML]{D9D9D9} 
CHAINS  & whisper & 1332     & 36     & Irish     & 3        & test  \\ \hline
\end{tabular}%
}
\end{table}
\section{Experimental Results}
\label{sec:exp}
According to the results from E2E systems based on conformers and transformers, which typically require substantial amounts of data, the utilization of pre-trained self-supervised models can be an effective strategy to address the data scarcity challenge in whisper ASR. As previously mentioned, conventional systems are inadequate for recognizing whispered speech. To validate this claim, a baseline system utilizing the OpenAI \textit{Whisper} model—one of the most powerful speech recognition systems for the English language—has been employed. No fine-tuning was performed on the \textit{Whisper} model for this work. We evaluated this system using whispered speech from the CHAINS dataset, with transcription performed using the beam search algorithm.

The ASR results using the Large-v2 version of the \textit{Whisper} model as the baseline system are presented in Table \ref{tab:t2}. This system achieved a WER of 16.69\% and CER of 3.81\% for normal and WER of 18.8\% and CER of 4.24\% for whispered speech parts of the CHAINS dataset. The speech texts in the dataset are identical for both speech styles. As shown, the system experienced a drop in performance when dealing with whispered speech. In addition, the performance of the Irish dialect was lower than that of the typical English \cite{radford2023robust}. These results indicate that whispered speech with a different dialect presents substantial modeling complexities for a system trained predominantly on normal speech data, underscoring the need for a dedicated system for whispered speech recognition.

In the subsequent step, the proposed system was trained using the Base+ version of WavLM model. This model comprises 94.7 million parameters and was pr-etrained over one million steps on 94,000 hours of speech data. The proposed system was fine-tuned using 17 hours of wTIMIT whispered and normal speech with a Singaporean dialect, along with 3 hours of normal speech with an Irish dialect from the CHAINS dataset. This training of WavLM was conducted using the PaddleSpeech toolbox \cite{zhang2022paddlespeech} on an NVIDIA A100 80GB GPU over 25 epochs with a batch size of 6 and an adaptive learning rate schedule. SpecAugment \cite{park2019specaugment} with time masking was employed as a data augmentation method, and the AdamW algorithm \cite{loshchilov2017decoupled} was used for optimization. Decoding was performed with a beam search size of 1 and a greedy search size of 16. The results of this system are presented in Table \ref{tab:t3}.
\begin{table}[t]
\caption{WER\% and CER\% of pre-trained \textit{Whisper} model for normal
and whispered speech styles}
\label{tab:t2}
\centering
\resizebox{0.9\linewidth}{!}{ % Adjust the width as needed
\renewcommand{\arraystretch}{1.5} % Increase the row height
\begin{tabular}{|>{\hspace{0.2cm}}l<{\hspace{0.2cm}}|>{\hspace{0.2cm}}l<{\hspace{0.2cm}}@{\hspace{0.4cm}}>{\hspace{0.2cm}}l<{\hspace{0.2cm}}|}
\hline
Metric & Normal Speech & Whispered Speech \\ \hline
WER\%       & 16.69       & 18.80        \\ 
CER\%       & 3.81        & 4.24          \\ \hline
\end{tabular}
}
\end{table}

\begin{table}[t]
\caption{WER\% and CER\% of fine-tuned WavLM model for whispered speech style with two decoding methods}
\label{tab:t3}
\centering
\resizebox{0.9\linewidth}{!}{ % Adjust the width as needed
\renewcommand{\arraystretch}{1.5} % Increase the row height
\begin{tabular}{|>{\hspace{0.2cm}}l<{\hspace{0.2cm}}|>{\hspace{0.2cm}}l<{\hspace{0.2cm}}@{\hspace{0.4cm}}>{\hspace{0.2cm}}l<{\hspace{0.2cm}}|}
\hline
Metric & Greedy Search & Beam Search \\ \hline
WER\%       & 9.28       & 9.22        \\ 
CER\%      & 2.60       & 2.59        \\ \hline
\end{tabular}
}
\end{table}

The proposed WavLM-based system achieved a WER of 9.28\% using the greedy search algorithm and 9.22\% based on the beam search algorithm, and the best achievement for the CER metric was 2.59\% for greedy search. These results demonstrate that the system effectively learned the acoustic parameters of whispered speech, significantly outperforming the OpenAI \textit{Whisper} model baseline system. This improvement can be attributed to including whispered speech data in the training phase. Additionally, the Irish dialect, which posed a challenge for the baseline system, was effectively addressed during training using normal speech data from the CHAINS dataset. The choice of the WavLM network proved advantageous due to its suitability for training on small datasets and its superior acoustic modeling capabilities.

\section{Conclusion}
\label{sec:concl}
In this article, we proposed a speech recognition system to handle the challenging conditions of whispered speech and the Irish dialect. By using training data that closely aligns with the acoustic characteristics of the test conditions, our system effectively modeled the complex nature of whispered speech. The results demonstrate the superiority of our system, achieving a Word Error Rate (WER) of 9.22\%, significantly outperforming the OpenAI \textit{Whisper} system, which served as our baseline. Based on our knowledge, this is the first work on whispered speech recognition on the CHAINS dataset. Furthermore, our findings show that the system delivers strong performance on the CHAINS dataset, particularly in the context of whispered speech recognition.

However, several limitations must be acknowledged. The primary constraint of this research is the scarcity of whispered speech datasets, especially those that include dialect variations. Although existing datasets like wTIMIT and CHAINS were used, they primarily focus on the English language and lack the diversity needed for robust, generalized performance across different dialects and speech conditions.

For future work, two key areas could be explored to enhance the performance and applicability of whispered speech recognition systems. First, utilizing other self-supervised models could offer insights into how different architectures handle whispered speech, potentially leading to further improvements in performance. Second, designing and developing a new, comprehensive dataset for whispered speech, particularly one that includes various dialects and languages, would be a critical step toward overcoming the dataset scarcity issue.

\bibliographystyle{IEEEtran}
\bibliography{bare_conf}

\end{document}